\documentclass[onecolumn,showpacs,aps,prc,nofootinbib,floatfix]{revtex4}

\usepackage{amsmath}
\usepackage{bm}
\usepackage{graphicx}

\usepackage{epsfig}
\newcommand{\beq}{\begin{equation}}
\newcommand{\eeq}{\end{equation}}
\newcommand{\bea}{\vspace{0.25cm}\begin{eqnarray}}
\newcommand{\eea}{\end{eqnarray}}

\newcommand{\ro}{\mbox{{\boldmath
$\rho$}}}

\newcommand{\nb}{\mbox{{\bf
n}}}

\newcommand{\rb}{\mbox{{\bf
r}}}

\newcommand{\bb}{{{\bf b}}}

\def\lsim{\mathrel{\rlap{\lower4pt\hbox{\hskip1pt$\sim$}}
    \raise1pt\hbox{$<$}}}         
\def\gsim{\mathrel{\rlap{\lower4pt\hbox{\hskip1pt$\sim$}}
    \raise1pt\hbox{$>$}}}         

\newcommand{\landau}{L.D.~Landau Institute for Theoretical Physics,
        GSP-1, 117940, Kosygina Str. 2, 117334 Moscow, Russia}

\begin{document}


\title{
  Effect of collective nuclear vibrations on initial anisotropies in
  central Pb+Pb collisions: A comparative study in Glauber and Trento models}
\date{\today}

\author{B.G.~Zakharov}\affiliation{\landau}

\begin{abstract}
We perform a comparison of the Glauber
  and Trento model predictions for the initial
  state azimuthal anisotropies $\epsilon_{2,3}\{2\}$
  in central Pb+Pb collisions at the LHC energies.
  We incorporate the quantum collective
  effects in $^{208}$Pb by sampling nucleon positions
  with filterings in the quadrupole and octupole moments.
  We use the data-driven filters adjusted to reproduce                                        the mean squared quadrupole and octupole moments extracted
  from the experimental
  strength functions.
  We find that the collective effects reduce considerably
the $\epsilon_2\{2\}/\epsilon_3\{2\}$ ratio.
The reduction is more pronounced in the Trento model.  
We obtain
$\epsilon_2\{2\}/\epsilon_3\{2\}\approx 0.75\div 0.79(0.68\div 0.71)$
  in the Glauber(Trento) models.
These values allow to resolve the $v_2$-to-$v_3$ puzzle.
  
\end{abstract}
%

\maketitle
\noindent {\bf 1}. {\bf Introduction}.
Successful hydrodynamic simulations of
soft hadron production in heavy ion collisions at RHIC and LHC
are widely interpreted as evidence of
the quark-gluon plasma (QGP) formation 
\cite{hydro1,hydro2,hydro3}.
The most convincing evidence for the hydrodynamic behavior of the QGP
is the azimuthal anisotropy of hadron distributions.
It is characterized
by the coefficients $v_n$ in the Fourier series in azimuthal angle
\beq
\frac{dN}{d\phi }=\frac{N}{2\pi}
  \left\{1+{\sum_{n=1}^\infty}2v_n\mathrm{cos}\left[n\left(\phi-\Psi_n\right)
  \right]\right\},
\label{eq:10}
\eeq
where $N$ is hadron yield in a certain $p_T$ and rapidity bin,
$\Psi_n$ are the event reaction plane angles.
In the absence of fluctuations of the initial QGP density/entropy,
at midrapidity ($y=0$) and nonzero impact parameter, $\bb$,
the hydrodynamic evolution generates the azimuthal dependence
of the hadron spectra due to larger pressure gradient
along the impact parameter vector
for the almond shaped overlap of the colliding nuclei.
Without fluctuations of the initial QGP density,
only even terms appear in (\ref{eq:10}).
In general, for fluctuating initial QGP density, the hydrodynamic evolution
leads to nonzero $v_n$ for all $n$.
For central ($b=0$) collisions of spherical nuclei the overlap region
is round, and the nonzero $v_n$ are entirely driven by fluctuations of
the initial QGP density.
It was found
\cite{vn_Ollit,Niemi,vn1} that for each event, $v_n$ for $n\le 3$,
at small centralities, can be approximately written as
\beq
v_n\approx k_n\epsilon_n\,,
\label{eq:20}
\eeq
where $\epsilon_n$ are the azimuthal anisotropies of the initial
QGP entropy density, $\rho_s(\ro)$,
in the transverse plane, defined as
\cite{Teaney_en,Ollitraut_en}
\beq
\epsilon_n=\frac{\left|\int d\ro \rho^n e^{in\phi}\rho_s(\ro)\right|}
{\int d\ro \rho^n\rho_s(\ro)}\,.
\label{eq:30}
\eeq
Here the vector $\ro$ is calculated in the 
tranverse c.m. frame, i.e., $\int d\ro \ro\rho_s(\ro)=0$. 
In the linear response approximation (\ref{eq:20}),
the linear scaling also holds for cumulants
$v_n\{2\}=\sqrt{\langle v_n^2\rangle}$ and
$\epsilon_n\{2\}=\sqrt{\langle \epsilon_n^2\rangle}$.

The initial entropy distribution in $AA$ collisions
(at the proper time $\tau_0\sim 0.5-1$ fm \cite{hydro1,hydro2,hydro3,TRAJ1})
cannot be calculated from first principles.
The most popular phenomenological models, used for 
calculation of the initial QGP entropy in the event-by-event
Monte Carlo (MC) hydrodynamical simulations of $AA$-collisions, are
the Glauber wounded-nucleon model \cite{MCGL,GLISS2} and the Trento model
\cite{TR1,TR2}. 
Both these models have been quite successful
in the even-by-event hydrodynamic simulations of $AA$ collisions
at RHIC and LHC energies. In particular, they can describe
experimental data on the flow coefficients $v_n$ at moderate centralities.
However, it turned out that the hydrodynamical
models fail to reproduce the ratio $v_2\{2\}/v_3\{2\}\approx 1$ observed in
the ultra-central ($c\to 0$)
2.76 TeV \cite{CMS_v23} and 5.02 TeV \cite{ALICE_v23}
Pb+Pb collisions (recently, $v_2\{2\}/v_3\{2\}\approx 1$ has also been
obtained for 5.36 TeV Pb+Pb collisions \cite{ALICE536}).
The point is that at small centrality the hydrodynamic models give
$k_2/k_3\sim 1.2-1.4$ \cite{Olli_k23,v23_Heinz,Olli_Xe,v23_Luzum,k23_Hira},
and calculations within the MC Glauber and Trento models with the
Woods-Saxon (WS) distribution give $\epsilon_2\{2\}$ close
to $\epsilon_3\{2\}$.
Then, using (\ref{eq:20}), one obtains that for the ultra-central Pb+Pb collisions
$v_2\{2\}/v_3\{2\}>1$, which is in contradiction with the LHC data (this is called
the $v_2$-to-$v_3$ puzzle).

There have been several unsuccessful attempts to explain the $v_2\{2\}/v_3\{2\}$
ratio in ultra-central Pb+Pb collisions.
They  include: modification of the initial conditions
\cite{v23_Heinz,XNW} (aimed to reduce the ratio $\epsilon_2\{2\}/\epsilon_3\{2\}$),
modification of the viscosity coefficients \cite{v23_Gale}, and the QGP equation of state of
\cite{v23_EoS} (aimed to reduce $k_2/k_3$).
The common feature of these analyses is
the use of the spherical  WS distribution, which is usually
used for the doubly-magic $^{208}$Pb nucleus.
However, analysis \cite{Bertsch}, performed within the generator-coordinate
extension of the Hartree-Fock-Bogoliubov method, shows that the ground state
of $^{208}$Pb may have a considerable octupole ($L=3$) pear shape deformation
(characterized by parameter $\beta_3$), since the
potential energy has minima at $|\beta_3|\sim 0.0375$.
One can expect that for ultra-central Pb+Pb collisions
the 3D octupole deformations of the colliding nuclei
should increase 2D coefficient $\epsilon_3\{2\}$,
and, consequently, reduce the ratio $\epsilon_2\{2\}/\epsilon_3\{2\}$.
The effect of the octupole deformations on $v_2\{2\}/v_3\{2\}$ ratio has been addressed
in \cite{v23_Luzum} within the Trento model.
It was found that, for reasonable values
of $\beta_3$ ($\lsim 0.04$), the octupole deformation of $^{208}$Pb
cannot
resolve the $v_2\{2\}/v_3\{2\}$ puzzle. In \cite{Xu}, within the Trento model,
it was shown that, for a reasonable
agreement with the LHC data on the $v_2\{2\}/v_3\{2\}$ ratio, it is necessary
to have $\sqrt{\langle \beta_3^2\rangle}\approx 0.15$.
However, one can show (see below) that the experimental data on the
octupole strength function of $^{208}$Pb  do not support such large values of
$\beta_3$.

The analyses \cite{v23_Heinz,v23_Gale,v23_EoS,v23_Luzum,XNW,Xu}, performed with
the WS distribution (spherical or with $\beta_3$ deformation),
ignore the quantum collective effects in $^{208}$Pb.
It is known that for heavy nuclei the $L=2,3$ collective
deformations
are closely related to the isoscalar giant quadrupole and octupole
resonances \cite{BM,Greiner} (for more recent reviews see \cite{Speth,Roca}).
In \cite{Z1,Z2} we have suggested to model in the MC Glauber calculations
the collective nuclear effects
in $^{208}$Pb  by an appropriate filtering of the nucleon positions
in the quadrupole and octupole
moments, which guarantees that the mean
squared quadrupole and octupole moments, $\langle Q_{2,3}^2\rangle$, of the colliding nuclei
are consistent with the ones extracted from the experimental
quadrupole and octupole strength functions for $^{208}$Pb
using the
energy weighted sum rule
(EWSR) \cite{EWSR}.
The results of \cite{Z1} demonstrate that the MC sampling with the ordinary
spherical WS distribution of $^{208}$Pb nucleus
overestimates $\langle Q_{2}^2\rangle$
 by a factor of $2.25$
(the fact that the WS distribution overestimates
 $\langle Q_{2}^2\rangle$
was first pointed out in \cite{Z_e2}).
In \cite{Z1}, we have shown that in the Glauber model the $Q_2^2$ filtering
results in $\approx 13$\% 
reduction of $\epsilon_2\{2\}$ (without change in $\epsilon_3\{2\}$), i.e., reduces 
the ratio $\epsilon_2\{2\}/\epsilon_3\{2\}$ by $\sim 13$\%.
In \cite{Z2} we have performed calculations including the filtering
in the octupole moment squared as well. From the point of view of the
$v_2$-to-$v_3$ problem, 
it is important that for
the octupole moment the situation is opposite to that for
the quadrupole moment: the octupole moment squared calculated
through the EWSR turns out to be larger than that
for the ordinary WS distribution.
It means that both the quadrupole and octupole filters
lead to reduction of the $\epsilon_2\{2\}/\epsilon_3\{2\}$ ratio
(which is necessary for resolving the $v_2\{2\}/v_3\{2\}$ puzzle).
The results of \cite{Z2} show that the reduction
of the $\epsilon_2\{2\}/\epsilon_3\{2\}$ ratio caused by the
quadrupole filtering
is stronger than that caused by the octupole filtering. 
The combined quadrupole and octupole filtering
reduces the $\epsilon_2\{2\}/\epsilon_3\{2\}$ ratio by $\approx 16\div 20$\%.
It allows to obtain a reasonable agreement with the data on
$v_2\{2\}/v_3\{2\}$ in ultra-central Pb+Pb collisions at the LHC energies.
It is important that the method of \cite{Z1,Z2} is fully data-driven.

In the light of the successful solution of the $v_2$-to-$v_3$ puzzle
in the Glauber model by incorporating the collective effects
in $^{208}$Pb \cite{Z1,Z2},
naturally, the question arises whether this mechanism also works
for the Trento model
\cite{TR1,TR2}.
To answer this question, in the present work we perform a comparative 
study of the anisotropies $\epsilon_{2,3}\{2\}$ in central Pb+Pb collisions
within the Trento and Glauber models.
As in \cite{Z1}, calculations are performed for Pb+Pb collisions 
at zero impact parameter $b=0$, i.e. at zero $b$-centrality defined
as $c_b(b)=\pi b^2/\sigma_{in}^{AA}$ \cite{centrality}.
Experimentally, the centrality is estimated  through charged particle
multiplicity $N_{ch}$ in a certain kinematic region ($N$-centrality,
$c_N$).
There is some mismatch between
the $b$- and $N$-centralities \cite{centrality}
(due to fluctuations of $N_{ch}$ at a given impact parameter).
However, comparison of the results for $\epsilon_{2,3}\{2\}$
at $c_b=0$ and of \cite{Z2} for $c_N\to 0$,
shows that the mismatch between the $c_b$- and $c_N$-centralities has
a very small effect on the ratio
$\epsilon_2\{2\}/\epsilon_3\{2\}$. Our calculations within the Glauber model
show that the fluctuations of the impact parameter
at $c_N\to 0$ increases the
ratio $\epsilon_2\{2\}/\epsilon_3\{2\}$ by only $\sim 2.5 $\%.
Since the effect of the $b$-fluctuations on $\epsilon_{n}$
is connected with variation of the geometry of the overlap region
at large scale (about the nucleus size), it is reasonable to expect that
for the Trento model
its magnitude should be approximately the same. 

  The plan of the paper is as follows. In Sec.~2 we 
  discuss the MC schemes for the entropy production.
  In Sec.~3 we briefly sketch the filtering in sampling of the nucleon
  positions.
  In Sec.~4 we present numerical results.
  We give conclusions in Sec.~5.\\

\noindent {\bf 2}. {\bf Outline of the MC schemes for entropy distribution}.
As in \cite{Z1,Z2}, we use the MC Glauber model in the form developed
in \cite{Z_gl1,Z_gl2}.
It allows to perform simulations
accounting for the meson-baryon Fock component of the nucleon.
In \cite{Z2} it was found that the effect of meson-baryon component
on the azimuthal coefficients $\epsilon_{2,3}\{2\}$ in ultra-central Pb+Pb
collisions is small. Therefore, in the present analysis we will
perform calculations without the
meson-baryon component.
In the Glauber model the
entropy is generated in the
sources corresponding to the wounded nucleons and to the hard binary
collisions. For each pair of colliding
nucleons, the probability of hard binary collision is suppressed by a
factor $\alpha$ \cite{KN}.
For A+B collisions, the midrapidity entropy density
in the transverse plane is written as a sum over
the wounded nucleon sources (in A and B) and the binary collision ones:   
\beq
\rho_s^G(\ro)=\frac{(1-\alpha)}{2}[S_w^A(\ro)+S_w^B(\ro)]+
\sum_{i=1}^{N_{b}} s_{b}(\ro-\ro'_{i})\,,
\label{eq:40}
\eeq
\beq
S_w^{A,B}(\ro)=\sum_{i=1}^{N_{w}^{A,B}} s_w(\ro-\ro_i^A)\,,
\label{eq:50}
\eeq
where $N_{w}^{A,B}$ are the numbers of the wounded nucleons in A and B,
the last term  (\ref{eq:40}) corresponds to the binary collisions
(we assume that $s_w=s_b$). We use a Gaussian
parametrization
\beq  
s_{w,b}(\ro)=s_0\exp{\left(-\ro^2/\sigma^2\right)}/\pi \sigma^2\,
\label{eq:60}
\eeq
with $s_0$ the total entropy of the source, and $\sigma$ the Gaussian width
of the source.  
We assume that
for each binary collision the center of the entropy source
is located in the middle between colliding nucleons.
We use the relation between the pseudorapidity charged
density and the entropy $dS/dy=CdN_{ch}/d\eta$ with
$C\approx 7.67$ \cite{BM-entropy}. It allows to describe the fluctuations
of $s_0$ directly in terms of fluctuations of
the charged multiplicity $n=s_0/C$.
We describe the fluctuations of $n$ by the Gamma distribution
\beq
\Gamma(n,\langle n\rangle)=
\left(\frac{n}{\langle n\rangle}\right)^{\kappa-1}
\frac{\kappa^\kappa\exp\left[-n\kappa/\langle n\rangle\right]}
{\langle n\rangle \Gamma(\kappa)}\,.
\label{eq:70}
\eeq
We use
$\langle n\rangle\approx 5.32$ and $\kappa\approx 0.52$
fitted to the experimental $pp$ mean charged multiplicity 
and its variance at $|\eta|<0.5$ for $\sqrt{s}=5.02$ TeV
in our MC Glauber analysis \cite{Z_gl3}   
of the LHC data on Pb+Pb and Xe+Xe collisions
(we use $\alpha\approx 0.13$, also obtained in \cite{Z_gl3}).

In the Trento model \cite{TR1,TR2}, for A+B collision the midrapidity entropy
density in the transverse plane
\beq
dS/dy\propto (T_A^p+T_B^p)^{1/p}\,,
\label{eq:80}
\eeq
where $T_{A,B}$ are the fluctuating profile functions (that are $\propto
S_w^{A,B}$).
For $p=0$ (\ref{eq:80}) gives
$dS/dy\propto (T_AT_B)^{1/2}$, which can mimic \cite{TR1,TR2}
approximately
predictions of the QCD inspired EKRT \cite{EKRT} and
IP-Glasma \cite{IPGl} models.
For $p=1$, the Trento model
formula (\ref{eq:80}) is similar to the contribution of the wounded nucleons
in the Glauber model (\ref{eq:40}). Analyses of experimental data on
$AA$ collisions at RHIC and LHC energies \cite{TR1,TR-hydro1}
support $p\approx 0$, and the width parameter of the Gamma distribution
$\kappa\approx 1.4$.
We use these parameters in our calculations.
For $p=0$, the Trento formula for the entropy density,
in terms of our functions $S_w^{A,B}$ (\ref{eq:50}),
reads
\beq
\rho_s^T(\ro)=a\Big[S_w^A(\ro)S_w^B(\ro)    \Big]^{1/2}\,.
\label{eq:90}
\eeq
In applications, the normalization parameter $a$ in (\ref{eq:90}) is fitted
to the data on the charged multiplicity. However, for calculation of
$\epsilon_{n}$ the value of $a$ is not important.\\

\noindent {\bf 3}.
{\bf Incorporation of the collective effects in MC sampling of $^{208}$Pb}.
In this section we briefly sketch the incorporation of the collective
nuclear effects in the sampling of $^{208}$Pb
(we refer the reader to \cite{Z1,Z2} for more details).
As in \cite{Z1,Z2},
we use  the spherical WS distribution
\beq
\rho_{A}(r)=\frac{\rho_0}{1+\exp[(r-R_A)/d]}\,
\label{eq:100}
\eeq
with $R_{A}=6.62$ fm, and $d=0.546$ fm \cite{PHOBOS}.
%
In the ground state, the mean squared $L$-multipole moment, $\langle Q_L^2\rangle$,
of a spherical nucleus  can be defined
quantum mechanically as
\beq
\langle Q_L^2\rangle=
\langle 0 |F^{+}_LF_L|0\rangle\,,
\label{eq:110}
\eeq
\beq
F_L=\sum_{i=1}^A r^L_iY_{Lm}(\nb_i)
\label{eq:120}
 \eeq
 with ${\nb}_i=\rb_i/|\rb_i|$.
For the uncorrelated spherical WS distribution (\ref{eq:100}), one can easily
obtain\footnote{We ignore
  a very small effect of the c.m. nucleon correlations. However,
  in our numerical simulations they have been treated properly.}
  \beq
  \langle Q_L^2\rangle_{WS}=
  \frac{A(2L+1)\langle
    r^{2L}\rangle}{4\pi}\,.
  \label{eq:130}
  \eeq
  Quantum mechanical calculation
of $\langle Q_L^2\rangle$,
  using the EWSR for the  strength function, $S_L(\omega)$, of the operator
  $F_L$, gives \cite{Z1,Z2}
\beq
\langle Q_L^2\rangle_{EWSR}=\langle Q_L^2\rangle_{WS}/r_L\,,\,\,\,\,\,\,\,
r_L=
\frac{ 2m_NE_c \langle r^{2L}\rangle}
     {L(2L+1)\langle r^{2L-2}\rangle}\,,
     \label{eq:140}
      \eeq
      where $m_N$ is the nucleon mass, and $E_c$ is the so called
      centroid energy. 
$E_c$ can be viewed as the typical excitation energy
for the operator $F_L$ acting on the ground state.
Using the Breit-Wigner parametrization of
the strength function
for $N$ resonances, one obtains \cite{Z2}
\beq
E_c=\left[\sum_{i=1}^{N}\frac{2f_i}{\pi\omega_{i}}\mbox{arctg}{(2\omega_i/\Gamma_i)}
  \right]^{-1}\,
\label{eq:150}
\eeq  
with $f_i$ the fraction of the $i$th resonance contribution to the EWSR.

For $L=2$, the EWSR is practically exhausted by the isoscalar giant quadrupole
resonance  with $\omega\approx 10.89$ MeV and
$\Gamma\approx 3$ MeV \cite{Youngblood_PRC69}. With these values one can
obtain from (\ref{eq:150}) $E_c\approx 11.9$ MeV \cite{Z2}, for which
(\ref{eq:130}) gives $r_2\approx 2.25$ \cite{Z2}.
The strength  function for $L=3$
is spread over a broad $\omega$ range.
It has several
peaks in the low-energy region 
$\omega\lsim 7$ MeV,
\cite{Harakeh_NPA327,Yamagata_PRC23,Fujita_LEOR1992},
and a broad resonance at 
$\omega\sim 16-20$ MeV with $\Gamma\sim 5-8$ Mev
\cite{Yamagata_PRC23,Pitthan_HEOR,Sasao,Carey_HEOR,Davis_HEOR,Youngblood_PRC69}.
Unfortunately, the experimental uncertainties in the EWSR fraction factors
for $L=3$ are large.
This does not allow for an accurate calculation of $r_3$.
Analysis \cite{Z2} shows that the available data on the $L=3$
strength function support the value of $r_3$  within the range of
$0.7-0.84$\footnote{
Note that in the scenario of \cite{Xu}
with the $\theta$-dependent WS distribution with $|\beta_3|\approx 0.15$,
$\langle Q_3^2 \rangle_{\beta_3=0.15}\approx
2\langle Q_3^2 \rangle_{WS}$. This is in contradiction
with $\langle Q_3^2 \rangle_{EWSR}$, that is smaller 
by a factor of $\sim 1.4\div 1.7$.}.

The obtained values of $r_{2,3}$ indicate that the standard MC
sampling with uncorrelated WS distribution overestimates the $L=2$ and
underestimate the $L=3$ fluctuations for $^{208}$Pb.
In \cite{Z1,Z2}, to cure this problem, 
it was suggested to perform the MC sampling of the nucleon positions with
the $Q_{2,3}^2$ filters that ensure that
$\langle Q_{2,3}^2\rangle=\langle Q_{2,3}^2\rangle_{EWSR}$.
For $L=2$, we need a $Q_{2}^2$ filter that pushes the distribution in $Q_{2}^2$
towards lower values of
$Q_{2}^2$\footnote{Note that, since $r_2>1$, the problem with the
  quadrupole moment
  cannot be eliminated by using the $\beta_2$ deformed WS distribution.}.
On the contrary, for $L=3$, we need
a $Q_{3}^2$ filter that pushes the distribution in $Q_3^2$
towards larger values of $Q_3^2$.
As in \cite{Z2}, we perform filtering in $Q_{2,3}^2$ by introducing
the suppression factors $\exp(-(Q_2^2/a_2)^2)$ for $L=2$, and
$[1-\exp(-(Q_3^2/a_3)^2)]$ for $L=3$ (with $a_{L}$ adjusted
to reproduce $\langle Q_L^2\rangle_{EWSR}$).

One might worry that
the theoretical predictions for $\epsilon_n$ could depend strongly on
the specific form of the used filters. To clarify this,  in \cite{Z2} we have
also performed calculations with sharp filters that select
only configurations with $Q_2^2<Q_{2\text{max}}^2$ and $Q_3^2>Q_{3\text{min}}^2$.
It was found that, although  the smooth and sharp filters lead to
radically different distributions  in $Q_2^2$ and $Q_3^2$,
the predictions for $\epsilon_{2,3}\{2\}$ obtained with these two methods are
very close to each other. We have checked that for the Trento model
the situation is similar. Thus, we see that both for the Glauber and Trento
models, the 2D azimuthal anisotropies $\epsilon_{2,3}\{2\}$ depend mostly on
$\langle Q_{2,3}^2\rangle$, and are practically insensitive to the
shape of the filters.
This conjecture is also corroborated by calculations with a filtering
excluding configurations with the internucleon distances smaller
than the $NN$ hard core radius $r_c\sim 1$ fm ($F(r_c)$ filter)
\cite{Bron_NNcore,GLISS2}.
In \cite{Z2}, we have found that in the Glauber model
predictions for $\epsilon_{2,3}\{2\}$
for the $Q_{2,3}^2$ filtering combined with the $F(r_c)$ filtering and
for the $Q_{2,3}^2$ filtering alone
are very close to each other (if in both cases the $Q_{2,3}^2$ filter
parameters adjusted to reproduce $ \langle Q_{2,3}^2\rangle_{EWSR}$).
We have checked that this holds for the Trento model as well.\\

\noindent {\bf 4}. {\bf Numerical results}.
We perform calculations of the root-mean-square anisotropies
$\epsilon_{2,3}\{2\}$ for 5.02 TeV Pb+Pb collisions
(in our previous study \cite{Z2}, we have found that predictions
for 2.76 and 5.02 TeV Pb+Pb collisions are practically the same).
Both for the Glauber and Trento models, the results
have been obtained by generating  $20000 $
Pb+Pb collisions at $b=0$.
The statistical errors in our results for
 $\epsilon_{2,3}\{2\}$ (estimated in the Glauber model) are
$\sim 0.2\div 0.3$\%.
We present results for the Gaussian width $\sigma=0.6$ fm.
We perform calculations for four versions of the sampling of the nucleon
positions:
(a) without filters, (b) with $Q_{2}^2$ filter,
(c) with $Q_{3}^2$ filter, (d) with $Q_{2,3}^2$ filters.
For the Glauber model the results are listed in Table I, and for
Trento model in Table II. We present
results for two versions of the $Q_3^2$ filters: for
$r_3=0.7$ and $r_3=0.84$ (numbers in brackets in columns (c) and (d)).

By comparing the results given in Tables I and II, we see
that for versions (a)-(d) $\epsilon_2^{Gl}\{2\}/\epsilon_2^{Tr}\{2\}\approx
1.22\div 1.3$, and $\epsilon_3^{Gl}\{2\}/\epsilon_3^{Tr}\{2\}\approx
1.15$.
From Tables I and II, we see that
the $Q_{2}^2$ filtering reduces 
$\epsilon_2\{2\}$ by $\approx 13(17)$\% for the Glauber(Trento) models,
and the $Q_{3}^2$ filtering increases
$\epsilon_3\{2\}$ by $\approx 8(10)$\%
at $r_3=0.7$ and by $\approx 3(3.5)$\% at $r_3=0.84$ for the Glauber(Trento)
models. Thus, we see that the effect of the $Q_{2}^2$ filtering
on $\epsilon_2\{2\}$ is more pronounced for the Trento model.
The effect of the $Q_{3}^2$ filtering on $\epsilon_3\{2\}$
is only slightly stronger for the Trento model. The average (for $r_3=0.7$ and
$0.84$) increase
in $\epsilon_3\{2\}$,  due to 
the $Q_{3}^2$ filtering, is $\approx 5.5(6.7)$\% for the Glauber(Trento) model.
It is by a factor of $\approx 2.4(2.5)$ smaller
than the reduction in $\epsilon_2\{2\}$ for the $Q_{2}^2$ filtering.
The combined $Q_{2}^2$ and $Q_{3}^2$ filterings reduce the 
ratio $\epsilon_2\{2\}/\epsilon_3\{2\}$ by $\approx 19(16)$\% for the Glauber
model, and by $\approx 24(21)$\% for the Trento model at $r_3=0.7(0.84)$.

To estimate $v_2\{2\}/v_3\{2\}$ from our predictions for the
$\epsilon_2\{2\}/\epsilon_3\{2\}$ ratio,
we use $k_2/k_3\approx 1.35$ extracted from the results for $v_{2,3}\{2\}$ and
$\epsilon_{2,3}\{2\}$ obtained with the TRAJECTUM code in \cite{TRAJ1}.
To account for the
mismatch between $b$ and $N_{ch}$ centralities (discussed in the Introduction),
we multiply our predictions for $\epsilon_2\{2\}/\epsilon_3\{2\}$
by a correction coefficient $\approx 1.025$ (obtained using the
results of \cite{Z2} at small $c_N$-centrality).
Using the results for 
$\epsilon_2\{2\}/\epsilon_3\{2\}$ shown in the last columns
(w/ $Q_{2,3}^2$ filters) of Tables I and II,
we obtain: $v_2\{2\}/v_3\{2\}\approx 1.05(1.1)$
for $r_3=0.7(0.84)$ in the Glauber model, and
$v_2\{2\}/v_3\{2\}\approx 0.943(0.983)$ for $r_3=0.7(0.84)$
in the Trento model.
These values are in reasonable
agreement with the CMS result $v_2\{2\}/v_3\{2\}\approx 0.996\pm 0.08$ \cite{CMS_v23}
for 2.76 TeV Pb+Pb collisions at $c_N\sim 0.002$\%,
and with the recent ALICE measurement $v_2\{2\}/v_3\{2\}\approx 1.045\pm 0.06$
\cite{ALICE536} for 5.36 TeV Pb+Pb collisions at $c_N\sim 0.0025$\%.

To understand the sensitivity of our predictions
to the value of
the Gaussian width $\sigma$ (in formula (\ref{eq:60})),
we also performed calculations for $\sigma=0.4$ and $0.8$ fm.
As in \cite{Z2} for small centralities, it was found that
the Glauber model predictions for $\epsilon_{2,3}\{2\}$ have low sensitivity
to the value of $\sigma$. For $\sigma=0.4$ fm,
$\epsilon_2\{2\}(\epsilon_3\{2\})$
increases by $\approx1.2(2)$\%.
For $\sigma=0.8$ fm, $\epsilon_2\{2\}(\epsilon_3\{2\})$
decreases by $\approx 2(3.5)$\%. The variation of
$\epsilon_2\{2\}/\epsilon_3\{2\}$ is $\lsim 1$\%.
For the Trento
model the variation of $\epsilon_{2,3}\{2\}$ with $\sigma$ is stronger:
for $\sigma=0.4(0.8)$ fm, both $\epsilon_2\{2\}$ and $\epsilon_3\{2\}$
increase(decrease) by $\approx$10.5(6)\%.
However, the ratio $\epsilon_2\{2\}/\epsilon_3\{2\}$ remains practically unchanged.
It is important that, both for the Glauber and Trento models, the variation
of $\epsilon_2\{2\}/\epsilon_3\{2\}$ with $\sigma$ turns out to be much smaller than
the effect of the $Q_{2,3}^2$ filterings.\\

\begin{table}
  \begin{tabular}{c|c|c|c|c}
\multicolumn{5}{c}{\mbox{Glauber model}} \\
\hline
Sampling & (a) w/o filters   & (b) w/ $Q^2_{2}$ filter
& (c) w/ $Q^2_{3}$ filter  & (d) w/ $Q^2_{2,3}$ filters \\
\hline
$\epsilon_2\{2\}$&0.111  &0.0964 &0.112(0.111)&0.0962(0.0967) \\
$\epsilon_3\{2\}$&0.119 &0.118  &0.128(0.123)&0.127(0.122) \\
$\epsilon_2\{2\}/\epsilon_3\{2\}$ &0.937 &0.816 &0.875(0.905)&0.756(0.792) \\
\hline
\end{tabular}
  \caption{
    The root-mean-square anisotropies $\epsilon_{2,3}\{2\}$
    and the $\epsilon_{2}\{2\}/\epsilon_{3}\{2\}$ ratio
    for central $5.02$ TeV Pb+Pb collisions
    obtained within the Glauber model. The $Q_2^2$ filtering
    is for $r_2=2.25$, the $Q_3^2$ filtering is
    for $r_3=0.7$ and $0.84$ (numbers in brackets) 
(see main text
    for details).}
\label{eps1}
\end{table}
\begin{table}
  \begin{tabular}{c|c|c|c|c}
\multicolumn{5}{c}{\mbox{Trento model}} \\
\hline
Sampling & (a) w/o filters   & (b) w/ $Q^2_{2}$ filter
& (c) w/ $Q^2_{3}$ filter  & (d) w/ $Q^2_{2,3}$ filters \\
\hline
$\epsilon_2\{2\}$&0.0903  &0.0753 &0.0902(0.0901)&0.0755(0.075) \\
$\epsilon_3\{2\}$&0.101 &0.102  &0.111(0.106)&0.111(0.106) \\
$\epsilon_2\{2\}/\epsilon_3\{2\}$ &0.893 &0.739 &0.811(0.847)&0.679(0.708) \\
\hline
\end{tabular}
\caption{
The same as in Table I, but for the Trento model.}  
\label{eps1}
\end{table}


\noindent {\bf 5. Summary}.
We have performed a comparative study of the influence of the collective
quantum effects in $^{208}$Pb
on the azimuthal anisotropies $\epsilon_{2,3}\{2\}$
in central Pb+Pb collisions at the LHC energies
for the MC Glauber and Trento models.
We account for the collective effects in colliding $^{208}$Pb nuclei
by sampling
the nucleon positions with filters in the squares of the quadrupole and
octupole moments ($Q_{2,3}^2$). The $Q_{2,3}^2$ filters have been adjusted
to guarantee that the mean
values $\langle Q_{2,3}^2\rangle$
coincide with the ones obtained using the EWSR from experimental data on the
quadrupole and octupole
strength functions of $^{208}$Pb. 
%
Our calculations show that, both for the Glauber and Trento models,
to good accuracy the $Q_n^2$ filtering affects only $\epsilon_n$. 
We have found that the effect of the $Q_{2}^2$ filtering on
$\epsilon_2\{2\}$ is stronger for the Trento model:
it reduces
$\epsilon_2\{2\}$ by $\approx 17$\% for the Trento model
and by $\approx 13$\% for the Glauber model.
The effect of the $Q_{3}^2$ filtering on  $\epsilon_3\{2\}$
is also stronger for the Trento model:
the $Q_{3}^2$ filtering increases
$\epsilon_3\{2\}$ by $\sim 3.5\div 10$\% for the Trento model, and 
by $\sim 3\div 8$\% for the Glauber model.
From the point of view of the $v_2$-to-$v_3$ puzzle, it is important
that both the $Q_2^2$ and $Q_3^2$ filterings reduce the  
$\epsilon_2\{2\}/\epsilon_3\{2\}$ ratio.
Our calculations predict that the combined $Q_{2,3}^2$ filtering
reduces $\epsilon_2\{2\}/\epsilon_3\{2\}$ by $\approx 16\div 19$\% for the Glauber
model, and by $\approx 21\div 24 $\% for the Trento model.
Both in the Glauber and Trento models,
the reduction of the $\epsilon_2\{2\}/\epsilon_3\{2\}$ ratio 
turns out to be strong enough (especially in the Trento model)
to resolve the $v_2$-to-$v_3$ puzzle. The main effect is due to
the filtering in the quadrupole moment.
It should be noted that, in contrast to analyses \cite{v23_Luzum,Xu},
in the present calculations the nuclear effects are accounted for
in a data-driven manner using
experimental data on the strength functions of
$^{208}$Pb, that allows to define uniquely 
the parameters of the $Q_{2,3}^2$ filters.
\\

\begin{acknowledgments}
I am grateful to S.P.~Kamerdzhiev  for helpful communications on physics
of the giant resonances.\\
\end{acknowledgments}

\end{document}